\documentclass[journal=langd5,manuscript=article,layout=twocolumn]{achemso}

\usepackage[version=3]{mhchem} 
\usepackage[T1]{fontenc}       
\usepackage{siunitx} 		   
\usepackage{xcolor}
\usepackage[normalem]{ulem}

\author{Guillaume Durey}
\affiliation
{Laboratoire Gulliver, UMR CNRS 7083, ESPCI Paris, Universit\'{e} PSL, 10 rue Vauquelin, 75005 Paris, France}
\altaffiliation
{School of Engineering, Brown University, 184 Hope Street, Providence, RI 02912, USA}
\author{Yoko Ishii}
\affiliation{Department of Physics, Graduate School of Science, Kyoto University, Oiwake-cho, Kitashirakawa, Sakyo-ku, Kyoto, 606-8562, Japan}
\author{Teresa Lopez-Leon}
\affiliation
{Laboratoire Gulliver, UMR CNRS 7083, ESPCI Paris, Universit\'{e} PSL, 10 rue Vauquelin, 75005 Paris, France}
\email{teresa.lopez-leon@espci.fr}

\title{Temperature-driven anchoring transitions\\at liquid crystal / water interfaces}

\abbreviations{LC, 5CB, PVA, SDS, OTS, BTBP, 6CB}

\begin{document}

\begin{tocentry}

\includegraphics{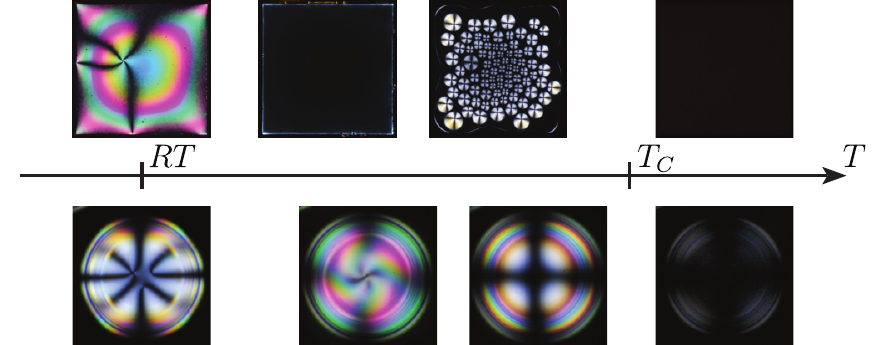}

\end{tocentry}


\begin{abstract}
Controlling the anchoring of liquid crystal molecules at an interface with a water solution influences the entire organization of the underlying liquid crystal phase, which is crucial for many applications. The simplest way to stabilize such interfaces is by fabricating droplets of liquid crystal in water; however, a greater sensitivity to interfacial effects can be achieved using liquid crystal shells, i.e. spherical films of liquid crystal suspended in water. Anchoring transitions on those systems are traditionally triggered by the adsorption of surfactant molecules onto the interface, which is neither an instantaneous nor a reversible process. In this study, we report the ability to change the anchoring of 4-cyano-4'-pentylbiphenyl (5CB), one of the most widely used liquid crystals, at the interface with dilute water solutions of polyvinyl alcohol (PVA), a polymer commonly used for stabilizing liquid crystal shells, simply by controlling the temperature in the close vicinity of the liquid crystal clearing point. A quasi-static increase in temperature triggers an instantaneous reorientation of the molecules from parallel to perpendicular to the interfaces, owing to the local disordering effect of PVA on 5CB, prior to the phase transition of the bulk 5CB. We study this anchoring transition on both flat suspended films and spherical shells of liquid crystals. Switching anchoring entails a series of structural transformations involving the formation of transient structures in which topological defects are stabilized. The type of defect structure depends on the topology of the film. This method has the ability to influence both interfaces of the film nearly at the same time, and can be applied to transform an initially polydisperse group of nematic shells into a monodisperse population of bivalent shells. 
\end{abstract}

\section{Introduction}

Nematic liquid crystals are a fascinating class of soft matter systems. They combine the solid-like property of long-range orientational order with the fluid-like ability to reorganize under very weak external stimuli. Liquid crystal molecules exhibit preferential orientation at interfaces, a phenomenon called anchoring. These boundary conditions are transmitted to the bulk molecules via the elasticity of the liquid crystal. Therefore, by tuning the anchoring at the interfaces, one can influence the organization of the bulk of a liquid crystal sample. Hence control over anchoring is one of the most valuable tools for investigating the fundamental principles that govern the organization of liquid crystals systems. It also lies at the core of many applications, in particular in display technologies, where liquid crystals are in contact with a solid substrate. Consequently, anchoring effects on a solid boundary have been extensively studied.\cite{Jerome_Anchoring, Cognard_Alignment} Two parameters govern the anchoring of a liquid crystal: the preferred molecular orientation at the interface (known as the easy axis) and the energetic penalty associated to violating this orientation (known as the anchoring strength). One speaks of planar anchoring when the easy axis lies in the plane of the substrate; homeotropic anchoring when it is perpendicular to it; or conical anchoring when the angle is between \SI{0}{\degree} and \SI{90}{\degree}. 

Interfaces between liquid crystals and other immiscible liquids, such as water, have recently attracted special attention. From a fundamental point of view, fluid interfaces have remarkable properties: being free from chemical or roughness defects, they are much more spatially homogeneous than solid boundaries. In terms of anchoring, they only prescribe the out-of-plane angular component of the preferred orientation: the in-plane component is degenerate, enabling the bulk liquid crystal a greater freedom of organization. In addition, they have recently shown great potential for building chemical or biological sensors\cite{Abbott_OpticalMethods, Abbott_InterfacesReview, Park_DetectionReview}. Indeed, microscopic events such as the adsorption of a protein at the liquid crystal / water interface can trigger a change in anchoring, which propagates a different orientation to the bulk molecules, in turn inducing a macroscopic change in the optical appearance of the sample. Most of these sensors take the form of flat nematic/water interfaces\cite{Abbott_Science, DePablo_Amyloid}; however, some recent studies have shown the advantages of curved interfaces. They can be obtained by confining liquid crystals in capillaries \cite{Jang_Capillaries, Khan_BackscatteringBiosensor}, droplets \cite{Abbott_Droplets, Abbott_DropletVsFilm, Kumar_Doplets}, or spherical shells \cite{Park_ShellSensor}. 

Nematic droplets are the simplest systems whose topology stabilizes defects in the liquid crystalline order -- points where the director, i.e. a nonpolar vector indicating the average molecular orientation at a mesoscopic scale, is undefined \cite{Lavrentovich_WordsAndWorlds, Teresa_Review}. An anchoring transition on a drop therefore translates to a modification of its topological constraints, leading to a striking change between well-known optical patterns. Yet, the low surface-to-volume ratio of a droplet makes its behavior mostly governed by bulk effects. Finally, liquid crystal shells -- spherical films of liquid crystal suspended in an aqueous solution -- provide a system that combines the sensitivity to interfacial events of thin films with the stabilization of topological defects of droplets\cite{Alberto_FirstShell, Teresa_FrustratedOrder, Teresa_Review, Lagerwall_Review}. The study of anchoring transitions in shells has a fundamental interest besides the development of applications. The two interfaces confining the shell allow for four different types of boundary conditions: planar-planar, planar-homeotropic, homeotropic-planar and homeotropic-homeotropic. Switching from one type of boundary condition to another usually triggers dynamical transitions involving the formation and recombination of topological defects \cite{Teresa_TopologicalTransformations, Teresa_DefectCoalescence, Lagerwall_InterfacesShells, Lisa_ChangeStripes, Lagerwall_PhototunableAnchoring, Lagerwall_KrafftTransition}. Transitions from purely planar or homeotropic states to hybrid states have been used to obtain information about the fundamental nature of topological defects\cite{Teresa_TopologicalTransformations, Lisa_ChangeStripes}, and to make them interact in unusual ways \cite{Teresa_DefectCoalescence}. In contrast, direct transitions between purely planar and purely homeotropic states remain elusive when using traditional methods to control the orientation of the liquid crystal molecules at fluid interfaces.

The traditional method for inducing anchoring transitions at liquid crystal / water interfaces, whether in flat films \cite{Abbott_First}, droplets\cite{Poulin_NovelColloidalInteractions, Poulin_NematicEmulsions}, or shells \cite{Teresa_TopologicalTransformations, Teresa_DefectCoalescence, Lagerwall_InterfacesShells, Park_ShellSensor, Lisa_ChangeStripes, Lagerwall_SurfactantsAnchoring, Lagerwall_KrafftTransition} consists in dissolving surfactant molecules in the water phase. Commonly used liquid crystals such as cyanobiphenyls display planar anchoring when in contact with water. The adsorption of surfactants onto the interface typically forces the liquid crystal molecules to reorient in the direction of the surfactant alkyl chains, therefore inducing homeotropic anchoring. This method comes with several shortcomings, especially when considering transitions on shells. Firstly, one cannot switch instantaneously between anchoring states in a controlled way. Indeed, the dynamics of the transition is governed by surfactant adsorption, whose spatiotemporal evolution cannot be controlled accurately. Secondly, the switching process is not reversible. While adding surfactants to the continuous phase without altering the shell position is tricky, flushing the shell with water to remove the surfactants without losing it is impracticable. Thirdly, one cannot induce an anchoring transition on both interfaces of a shell at the same time, as the outer interface is the only one where the concentration of surfactants can be tuned directly. Indeed, although these molecules can diffuse through the liquid crystal film, from the continuous aqueous phase to the inner water droplet, it typically takes them about a day to do so -- leaving the shell with ill-defined boundary conditions for an extended period of time. Despite recent efforts to induce controlled anchoring transitions in liquid crystal shells\cite{Lagerwall_KrafftTransition, Lagerwall_BlockCopolymers}, a powerful alternative to traditional surfactant-induced techniques is still missing.

\begin{figure*}[t!]
\centerline{\includegraphics{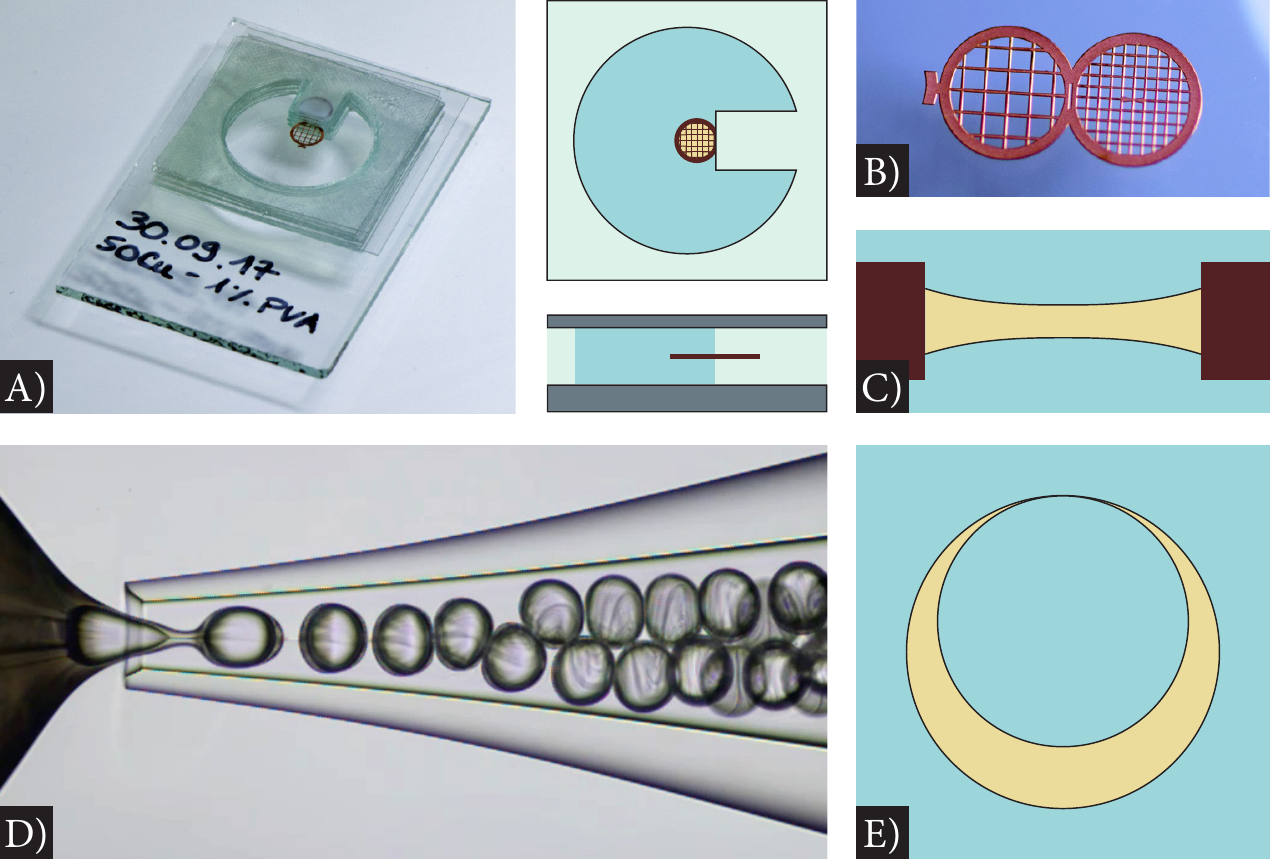}}
\caption{\label{MaterialsAndMethods} \textbf{Films and shells of nematic liquid crystals in water.} In this study, two complementary experimental systems were used: thin nematic films (top row) and nematic shells (bottom row). Nematic films were produced by fixing a copper double TEM grid in mid-height of an observation chamber filled with a water solution. One half of the double grid was embedded into the walls of the chamber, made of double-sided tape, while the other single grid was filled with a sub-microliter volume of liquid crystal. A) Photograph (left) and schematic representations (top right: top view, bottom right: side view) of the experimental chamber mounted on a microscope glass slide. B) Close-up photograph of  a double TEM grid. Each grid is \SI{3.05}{\mm} in diameter. C) Cross-section of a cell of the grid filled with liquid crystal. The cell is \SI{430}{\um} in width, \SI{18}{\um} in thickness. The film is about \SI{5}{\um} in the center. The vertical scale has been exaggerated ten times with respect to the horizontal scale. Nematic shells were produced by using a microfluidic device made of two tapered glass capillaries facing opposite directions, nested in a larger square capillary. D) Microphotograph of shell production. The tip of the capillary is about \SI{150}{\um} wide. E) Schematic representation of a vertical cross-section of a nematic shell. The thickness is heterogeneous due to a slight difference in density between the water phases and the 5CB. The typical outer diameter is \SI{100}{\um} while the typical mean thickness is \SI{10}{\um}.}
\end{figure*}

In this article, we report a new method to induce a change in boundary conditions at liquid crystal / water interfaces that addresses those issues. It relies on coating the interface with polyvinyl alcohol (PVA), a polymer widely used for stabilizing double emulsions and for enforcing strong planar anchoring. We show that PVA enables a richer control over the molecular anchoring at a fluid interface than what was previously thought: it is actually possible to tune its influence on the underlying liquid crystal phase through a precise control of the temperature of the system. We focus on the interface between water solutions of PVA and 4-cyano-4'-pentylbiphenyl (5CB), one of the most widely used liquid crystals. We demonstrate that the usual planar alignment at room temperature is replaced by perpendicular alignment to the interfaces when the temperature is brought just a few tenths of degrees Celsius below the clearing point of the liquid crystal. With this easily controllable experimental parameter, one can therefore have the system switch between two antagonist anchoring states. This anchoring transition is nearly instantaneous, as well as spatially homogeneous, and affects nearly simultaneously all the interfaces of the liquid crystal. We first study the transition in flat suspended nematic films, which provide a new experimental platform extremely sensitive to interfacial phenomena. We then demonstrate its usefulness by applying it on liquid crystal shells, in which the change of boundary conditions triggers a series of structural transformations, imposing the formation of topological defects with integer topological charge.

\section{Materials and methods}

For this study, we consider two different geometries: spherical nematic shells and flat suspended nematic films. Shells provide a simple way to stabilize a liquid crystal film in water without the assistance of any solid boundary. While surface tension tends to preserve such a spherical geometry, it makes flat films unstable. Flat suspended films are more challenging to stabilize, but provide a simpler geometry to interpret anchoring transitions in a straightforward way, since curvature and topological constraints are not present. It thus constitutes a good starting point for our study. Here we propose a strategy to stabilize such films.

A commonly used method for stabilizing water / liquid crystal interfaces, developed by Abbott and coworkers\cite{Abbott_First}, relies on confining the liquid crystal to a transmission electron microscopy (TEM) grid -- a small metal disk \SI{3.05}{\mm} in diameter and \SI{18}{\um} in thickness, having an array of holes with a given geometric shape. In Abbott's system, the grid is placed on top of a glass substrate, so that the pores of the grid become micro-sized observation cells. Confining the liquid crystal to these cells prevents it from dewetting the water interface by pinning it to a solid substrate. When the cells are immersed in an aqueous solution, the liquid crystal is sandwiched between a fluid and a solid interface. The glass solid substrate is usually coated with octadecyltrichlorosilane (OTS) in order to enforce homeotropic anchoring. These strong anchoring cues on the bottom interface constrain the system, reducing its sensitivity to subtle anchoring changes occurring at the free interface. 

\begin{figure*}[t!]
\centerline{\includegraphics{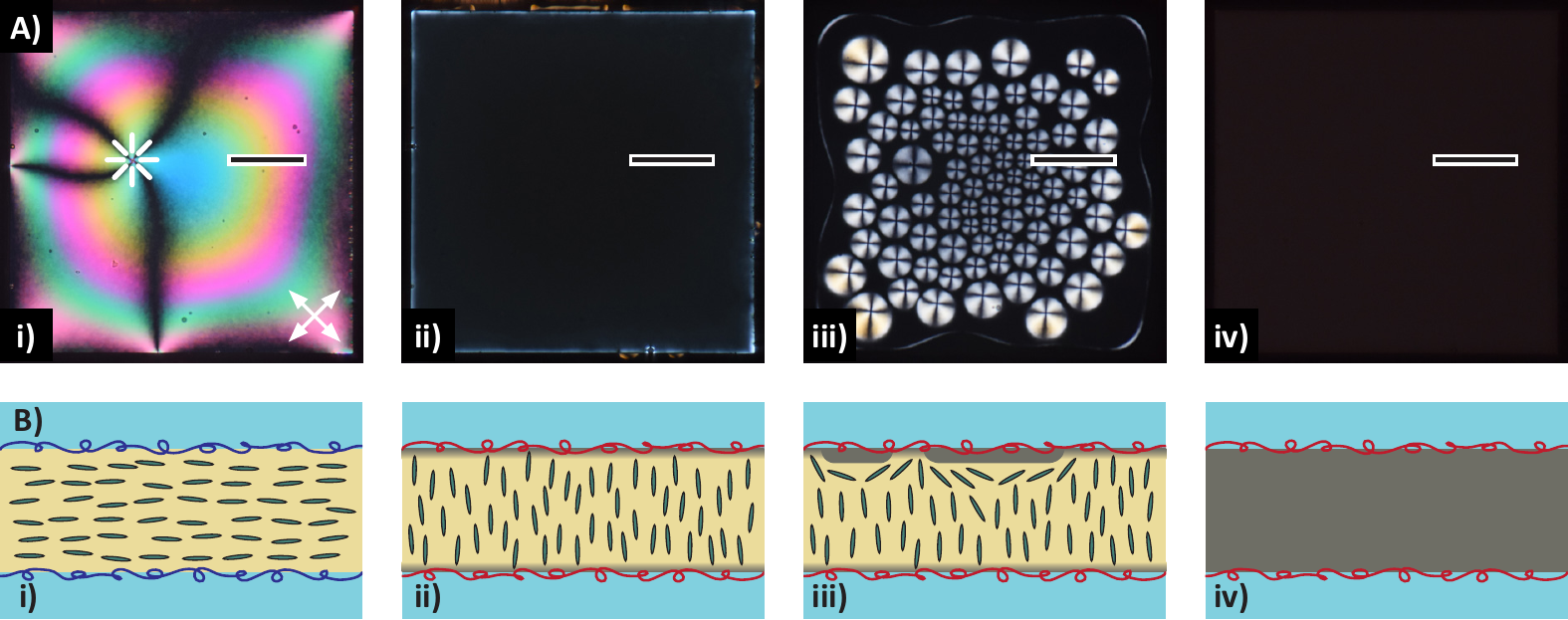}}
\caption{\label{TransitionFilmPVA} \textbf{5CB film in a \SI[detect-weight]{0.1}{\percent} wt PVA solution undergoing a slow temperature ramp} Top: series of top-view micrographs through crossed polarizers. Bottom: series of vertical cross-sections schematically depicting the molecular arrangement at the location of the black and white rectangle. Throughout, the film is \SI{430}{\um} wide, about \SI{10}{\um} thick close to the edges, about \SI{5}{\um} thick in the center, and is attached to a suspended copper TEM grid. The film undergoes a temperature ramp with a rate of \SI{0.01}{\degreeCelsius\per\minute}. i) In most of the nematic range, the film displays strong birefringence colors, an indication of the planar anchoring that aligns the molecules parallel to the interfaces. This particular film includes a topological defect; director field lines are sketched in its vicinity. The vertical cross-section is taken far from the defect, along a direction where the director is in the plane of the cross-section and uniform in this initial stage. ii) A few tenths of degrees Celsius below the phase transition temperature of pure 5CB, a molecular reorientation perpendicularly to the interfaces occurs, evidenced by a dark film with bright edges, likely due to the interfacial PVA-5CB mixture becoming isotropic. iii) When the temperature reaches the clearing point of pure 5CB, islands of bulk isotropic phase nucleate from the top and bottom interfaces of the film, inducing conical anchoring of the remaining nematic pure 5CB in the bulk. iv) The isotropic islands grow in size until the whole film turns isotropic. These four micrographs, taken respectively at \SI{-0.05}{\degreeCelsius}, \SI{-0.01}{\degreeCelsius}, \SI{+0.02}{\degreeCelsius} and \SI{+0.11}{\degreeCelsius} of the apparition of the first island of bulk isotropic phase, are frames of the timelapse provided in Supplementary Movie 1.}
\end{figure*}

Here, we propose a novel adaptation of Abbott's system, which addresses those issues. Indeed, we produced suspended films in water, in which the anchoring at the two interfaces is solely controlled by the aqueous solution. To do this, we used double "oyster" grids\cite{Jakli_Oyster}, consisting of two laterally-fused standard TEM grids. Fig. \ref{MaterialsAndMethods}A shows the observation chamber in which they were placed, and Fig. \ref{MaterialsAndMethods}B one of these grids. A \SI{500}{\um} high cylindrical observation chamber was built by stacking ten layers of double-sided tape (Gudy 804, Filmolux) on a glass slide. A double oyster grid (GD50/100-Cu, Gilder) was deposited on the tape, in such a way that one of the single grids was stuck to the tape while the second one was suspended in mid-air. Ten more layers of tape were added on top of the first ones, bringing the suspended grid in the middle plane of the one-millimeter-high chamber. \SI{0.1}{\micro\liter} of 5CB (Synthon Chemicals) or 6CB (Sigma-Aldrich) were deposited on the grid. The observation chamber was then filled with the aqueous phase, which could either be pure deionized water (Millipore Synergy), or a PVA solution (13-\SI{23}{\kilo\gram\per\mole}, 87-89\% hydrolyzed, Sigma-Aldrich). Filling the chamber also detached an excess of liquid crystal from overfilled cells of the grid, generating a few microdroplets which adsorbed to the hydrophobic walls of the chamber. The chamber was finally sealed with a glass cover slip. For observation, we focused on a single square hole in the grid, a vertical cross-section of which is sketched in Fig. \ref{MaterialsAndMethods}C. 

Nematic shells of 5CB were produced using a glass capillary device\cite{Utada_DoubleEmulsions}. In this type of device, a water / liquid crystal compound jet is sheared by a water-based outer solution using a combination of co-flow and flow-focusing geometries, and breaks up into double emulsions, as shown in Fig.~\ref{MaterialsAndMethods}C. To reach production regimes at reasonable flow rates, the viscosity of the outer aqueous phase was increased by adding \SI{60}{\percent} wt glycerol (VWR Chemicals). The inner aqueous phase was either pure deionized water or a \SI{1}{\percent} wt PVA solution. The shells were collected in a \SI{200}{\um} high cylindrical observation chamber built by stacking two layers of double-sided tape. The continuous collection phase was identical to the inner aqueous phase, with traces of glycerol originating from the outer aqueous phase in the capillary device. Due to a slight density difference between 5CB and the aqueous phases, the inner aqueous droplet typically rises to the top of the shell, making the shell heterogeneous in thickness.

We achieved an extremely fine level of temperature control in order to witness temperature-driven anchoring transitions on these experimental systems. The observation chambers were inserted into a TS62 thermal stage driven by a mK1000 unit, both from Instec. They were subjected to a \SI{0.01}{\degreeCelsius\per\minute} temperature ramp close to the clearing point of the liquid crystal (around \SI{34.5}{\degreeCelsius} for 5CB, around \SI{29.5}{\degreeCelsius} for 6CB). Temperature was controlled with an accuracy better than a tenth of degree Celsius. The sample's evolution was recorded using a Nikon DSLR camera (D5300 or D300s) mounted on an upright polarized optical microscope (Nikon Ni-U).

\section{Results and discussion}

\subsection{Temperature-driven anchoring\\transitions in nematic films}

\subsubsection{Parallel state}

As widely reported, 5CB displays planar anchoring when in contact with an aqueous solution containing PVA. This is consistent with the strong birefringence colors displayed by our suspended nematic films at room temperature, as shown in Fig.~\ref{TransitionFilmPVA}-i, which is a cross-polarised image of a 5CB film suspended in a \SI{0.1}{\percent} wt PVA solution. The colors become more and more vivid towards the center of the cell, indicating that the film is concave. Using a Michel-L\'{e}vy chart\cite{Sorensen_MichelLevy}, we estimate that its thickness ranges from \SI{10}{\um} on the edges to \SI{5}{\um} in the center. Besides, since the contact with the metallic substrate provides homeotropic anchoring\cite{Abbott_InterfacesReview}, the liquid crystal molecules are arranged perpendicularly to the edges of the film and parallel to the interfaces with water, yielding a radial defect in the center of the cell in the specific case of the film featured in Fig.~\ref{TransitionFilmPVA}. Defects can be characterized by their topological charge, which quantifies the number of times the director rotates around the singularity. The defect in our film has a [+1] topological charge, indicating a 2$\pi$-rotation of the director around the defect core. This can be distinguished by the number of dark brushes stemming from the defect, which is equal to four times the absolute value of its topological charge. 
 
\begin{figure*}[t!]
\centerline{\includegraphics{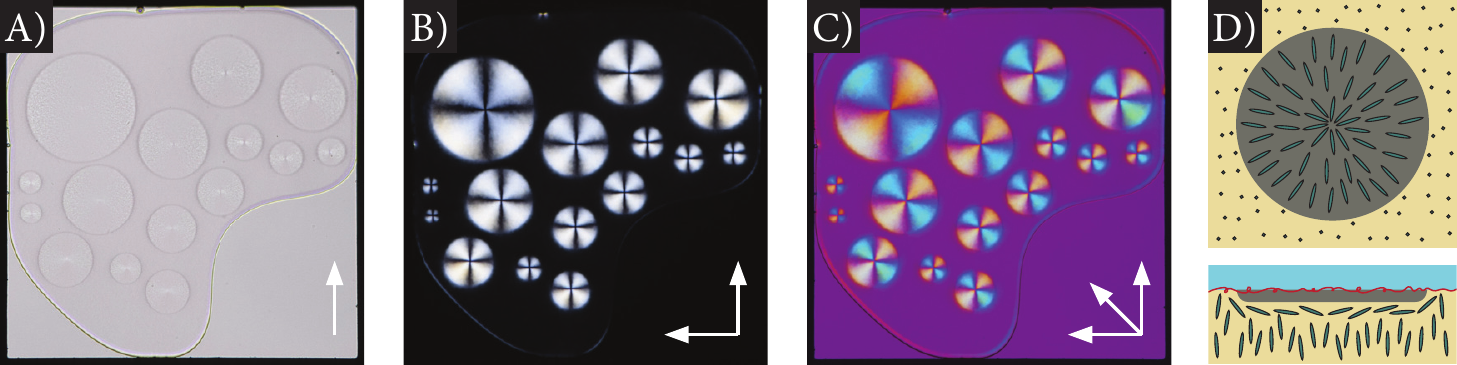}}
\caption{\label{ConicalIslands} \textbf{Coexistence of regions of pure isotropic phase embedded into a homeotropic nematic domain in the bulk of a 5CB film, close to the clearing point.} A) Bright field (single polarizer). From the edges to the center of the cell, one can make out the isotropic domain, the homeotropic nematic domain, and the circular islands with their central point defect. B) Crossed polarizers. The [+1] charge of all defects is revealed. First-order birefringence colors indicate that only a small thickness of liquid crystal has tilted back towards the planes of the interfaces. C) Crossed polarizers with first-order retardation plate. The radial orientation of the long axis of the molecules is revealed. Throughout, the film is \SI{430}{\um} wide. D) Schematic of a typical isotropic droplet, top and side view.}
\end{figure*}

Interestingly, we observe that the molecular anchoring at the interfaces with the aqueous solution can be changed by accurately controlling the film temperature in the vicinity of its clearing point. By quasi-statically increasing the temperature of the system with a linear rate of \SI{0.01}{\degreeCelsius\per\minute}, we were able to sweep sequentially through a series of distinct states with different boundary conditions until the entire sample turns isotropic. This evolution, recorded in Supplementary Movie 1, is as follows: the usual parallel state (Fig.~\ref{TransitionFilmPVA}-i), a new perpendicular state (Fig.~\ref{TransitionFilmPVA}-ii) and a coexistence state (Fig.~\ref{TransitionFilmPVA}-iii), before the isotropic state (Fig.~\ref{TransitionFilmPVA}-iv). 

\subsubsection{Perpendicular state}

A few tenths of degrees Celsius below the phase transition temperature of the bulk 5CB, an unexpected abrupt reorientation to a perpendicular state occurs. Indeed, the appearance of the cell changes abruptly to a black film with a bright rim (Fig.~\ref{TransitionFilmPVA}-ii), a pattern which is recognized as the signature of homeotropic anchoring at a liquid crystal / water interface\cite{Abbott_First}. The loss of birefringence is explained by the reorientation of the long axis of the molecules in the bulk perpendicularly to the interfaces: the optic axis of the liquid crystal becomes aligned with the optical axis of the microscope. The bright rim corresponds to the region where the director transitions to accommodate the homeotropic anchoring cues on the edges of the grid with the homeotropic anchoring cues on the top and bottom interfaces. We were able to maintain the film in this state for more than eight hours by keeping the temperature constant. 

The perpendicular orientation of the liquid crystal molecules was confirmed by studying 5CB films doped with an anisotropic fluorescent dye. We used the anisotropic dye N,N'-Bis(2,5-di-tert-butylphenyl)-3,4,9,10-perylenedicarboximide (BTBP), which aligns with the 5CB molecules \cite{Smalyukh_FastFCPM}. The intensity of the fluorescence signal is minimum when the dye molecules are oriented along the direction of light propagation. When imaging the transition from the birefringence colors to the dark state, we observed that the fluorescence in the homeotropic region was indeed less intense than in the planar region, as shown in Fig.~S1. Also as expected, the isotropic state at the end of the evolution of the film went back up to intermediate intensities, about two-thirds of the way from homeotropic to planar levels. We considered the fluorescence emission as being temperature-independent over such a small temperature interval. Besides, if the temperature were to influence the fluorescence intensity, we would expect a continuous decrease. Yet we report non-monotonic behavior.

\subsubsection{Coexistence state}

For a temperature even closer to the clearing point, the film enters a state of coexistence between the bulk nematic and isotropic phases. When the temperature is increased from the perpendicular state, the isotropic phase nucleates from the grid walls, creating a homeotropically-aligned nematic domain no longer in contact with the walls, as shown in Fig.~\ref{TransitionFilmPVA}-iii. The most striking feature of this state is the reappearance of birefringence in the form of a number of bright disk-like islands embedded in the homeotropic region, introduced in Fig.~\ref{TransitionFilmPVA}-iii and studied in detail in Fig.~\ref{ConicalIslands}. As the isotropic phase grows, the homeotropic domain shrinks -- in Fig.~\ref{ConicalIslands}, the homeotropic domain has shrunk more than in Fig.~\ref{TransitionFilmPVA}-iii. It eventually breaks up into multiple smaller domains, before the whole film turns isotropic.  When the temperature is lowered from the isotropic state, at the same rate of \SI{-0.01}{\degreeCelsius\per\minute}, homeotropically-aligned domains of nematic phase nucleate from the isotropic phase, then merge and grow until the whole film is homeotropic. Contrary to what happens upon heating the film, there are no birefringent islands formed at the coexistence state when the film is cooled. 

The structure of the islands can be revealed by investigating their properties, both static and dynamic (Fig.~\ref{ConicalIslands}, Supplementary Movie 1 and Fig.~S2). They stem from the nucleation of flattened regions of isotropic phase at the liquid crystal / water interface, extending into the bulk of the nematic phase. This situation, where pure isotropic 5CB coexists with pure nematic 5CB, has been described in the literature. \cite{Faetti_IsotropicAnchoring, Faetti_IsotropicAnchoringReview} The isotropic phase of 5CB was reported to enforce out-of-plane anchoring of its nematic phase, at an angle of \SI{26.5}{\degree} $\pm$ \SI{6}{\degree} from the surface. \cite{Faetti_IsotropicAnchoring} This explains the re-emergence of birefringence around the islands of isotropic 5CB. In bright field, all islands exhibit a point defect at their center and a sharp boundary with the surrounding homeotropic domain (Fig.~\ref{ConicalIslands}A). Under crossed polarizers, they display first-order blue-gray interference colors, meaning that the liquid crystal region contributing to the birefringence is very thin (Fig.~\ref{ConicalIslands}B). The defects are located at the intersection of four dark brushes which follow both polarizers when rotated, indicating that their topological charge is [+1]. When a first-order retardation plate is added, the liquid crystal molecules are found to be arranged radially from the defect (Fig.~\ref{ConicalIslands}C). The absence of [+1/2] defects -- which induce a $\pi$-rotation of the director, and thus, are only allowed when the boundary conditions are planar --  is consistent with the existence of conical anchoring cues at the boundaries\cite{Kim_IsotropicDroplets, Abbott_oligomers}. 

\begin{figure*}[ht]
\centerline{\includegraphics{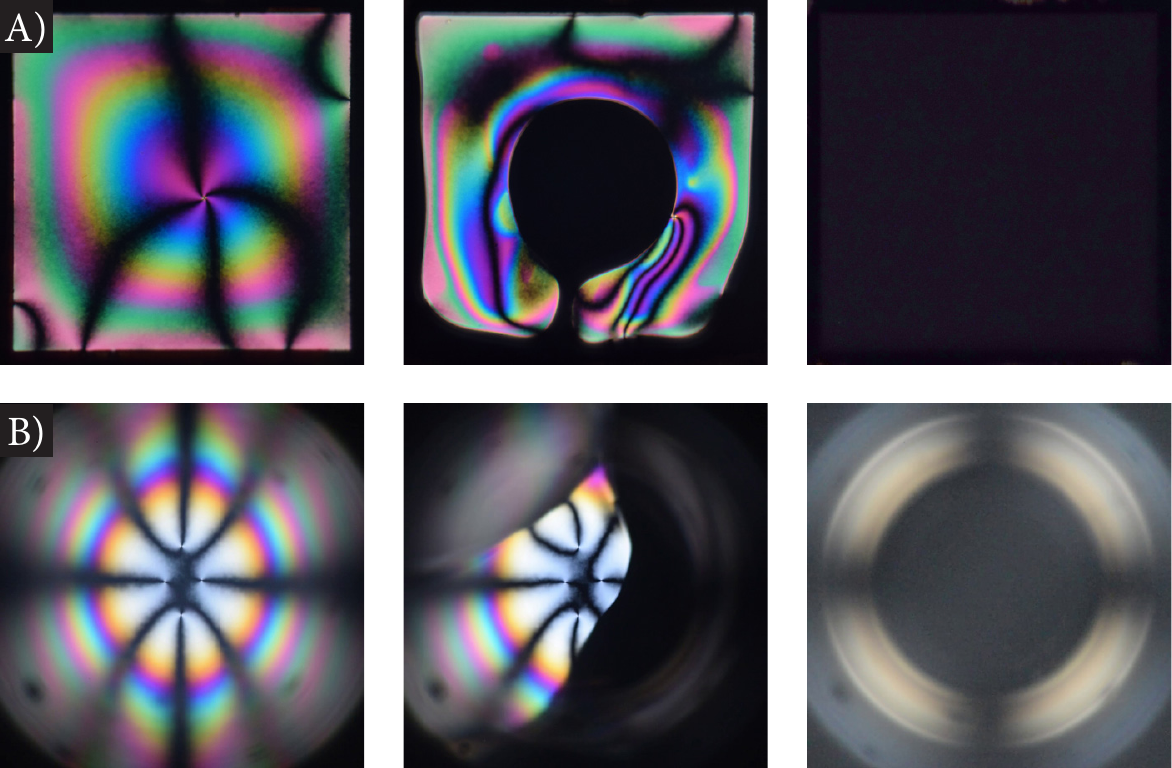}}
\caption{\label{TransitionsWithoutPVA} \textbf{5CB film and shell in pure water solutions undergoing a slow temperature ramp} As temperature of a film (A) or a shell (B) is increased, the nematic phase is progressively replaced by the isotropic phase, without any significant change in the shrinking planar texture. Throughout both experiments, the temperature ramp has a rate of \SI{0.01}{\degreeCelsius\per\minute}. The film is \SI{430}{\um} wide and the shell is about \SI{200}{\um} in outer diameter. The first series of micrographs corresponds to frames of the timelapse provided in Supplementary Movie 3, and the second series to frames of the real-time film provided in Supplementary Movie 4.}
\end{figure*}

Monitoring the evolution of the islands in time reveals additional information about their nature. The islands nucleate in the homeotropic domain. As temperature rises, they slowly grow in diameter, their color remaining unchanged. Then they meet one of two possible outcomes, or a combination of both. On the one hand, they can disappear by coalescence: two islands fuse from the side, creating a [-1] defect at the junction, which quickly annihilates with one of the [+1] defects from the original islands (Fig.~S2A). Yet islands can sometimes be seen to overlap without coalescing. In these instances, the interference colors add up but the geometry of the patterns remains unperturbed (Fig.~S2B). This demonstrates that while the islands are located within the bulk of the sample, they do not extend far from the liquid crystal / water interfaces in terms of depth. In most experimental realizations, they only form at either the bottom or the top interface, but sometimes they appear on both, in which case they can only interact with islands on the same interface. On the other hand, they can collide with the boundary between the homeotropic domain and the isotropic phase. In this case, they fuse with the isotropic phase (Fig.~S2A). We were able to trigger the formation of identical islands in a thin nematic film (approximately \SI{30}{\um} thick) confined between two glass plates treated to induce strong homeotropic anchoring undergoing a phase transition (see Supplementary Information and Supplementary Movie 2). This confirmed that their presence was indicative of a simple phase coexistence in the bulk between pure nematic and isotropic 5CB. Finally, we note that the same type of islands were reported by Abbott et al. very recently \cite{Abbott_oligomers}, and that their analysis is consistent with ours.

\subsection{The role of PVA: experimental clues and mechanistic insights}

Upon repeating these experiments for PVA solutions with decreasing concentrations, we were able to bring out the fact that a PVA concentration above a certain threshold was a necessary requirement for the temperature-driven anchoring transition to take place. Indeed, no anchoring transitions can be observed for samples with PVA below \SI{0.1}{\percent} wt in the aqueous solution. Films go directly from planar nematic to isotropic, as can be seen in Supplementary Movie 3 and Fig.~\ref{TransitionsWithoutPVA}A, where the aqueous solution is simply pure water. 

We were also able to prove that the value of the anchoring transition temperature was linked to the liquid crystal -- not to the PVA -- by conducting additional experiments with a different liquid crystal of the same series. While the clearing point of 5CB is around \SI{34.5}{\degreeCelsius}, 4-cyano-4'-hexylbiphenyl (6CB) has a phase transition temperature at around \SI{29.5}{\degreeCelsius}. Performing the same experiments with 6CB yielded identical results, albeit uniformly shifted in temperature: an anchoring transition was observed a few tenths of degrees Celsius below the clearing point when the liquid crystal was in contact with a \SI{0.1}{\percent} wt PVA solution; no anchoring transition was observed prior to the phase transition when in contact with pure water.

In the rest of this subsection, we offer suggestions into the nature of the mechanism at play in our experiments, although a conclusive experimental validation is out of the scope of the present paper.

From the experiments mentioned above, it seems that the temperature-induced anchoring transition observed in the presence of PVA stems from a change of interfacial order in the film. Since the PVA chains lie at the liquid crystal / water interface, they disrupt the nematic arrangement of the liquid crystal molecules in the vicinity. Therefore, the order parameter of the nematic phase is expected to be lower near the interface than in the bulk. With greater disorder comes a lower phase transition temperature: hence, such an interfacial mixture of PVA chains and 5CB molecules would have a slightly lower clearing point than the pure 5CB bulk. This suggests that our system undergoes an interfacial melting of the PVA-5CB mixture before the bulk pure 5CB turns isotropic. Thus, we hypothesize that once this interfacial layer melts, a new interface appears in our films: it separates the isotropic PVA-5CB surface layer from the bulk nematic phase of pure 5CB. Such a localized phase transition in the presence of adsorbates has been reported in recent studies. \cite{dePablo_AmphiphilePhaseTransition, Abbott_oligomers}

How does the presumed existence of this new interface cause the reorientation of the underlying liquid crystal molecules? The most probable hypothesis is that it induces a change in the easy axis accompanied by a decrease in anchoring strength. Indeed, it seems reasonable to think that the easy axis prescribed by the isotropic PVA-5CB surface layer is close to the one reported in the literature in the case of pure isotropic 5CB, \cite{Faetti_IsotropicAnchoring, Faetti_IsotropicAnchoringReview} thus imposing out-of-plane anchoring to the underlying bulk nematic film. Such boundary conditions would generate splay and bend distortions of the director field in the nematic phase. Yet, as the anchoring strength between the nematic and isotropic phases of a single compound is relatively weak, it may be energetically favorable for the system to relax these distortions immediately, through the uniform reorientation of the molecules perpendicularly to the interfaces, violating the new easy axis in the process. A similar mechanism has been used to explain recent experimental findings. \cite{dePablo_AmphiphilePhaseTransition, Abbott_oligomers, Lagerwall_BlockCopolymers} However, as far as we know, the easy axis at the interface of a bulk nematic phase with an isotropic PVA/5CB mixture has not yet been measured. An alternative possibility to the indirect mechanism above would be that the interfacial layer directly enforces homeotropic anchoring. Quantitative in-depth experiments are required to tell the two mechanisms apart.

\subsection{Application to nematic shells}

\begin{figure*}[ht]
\centerline{\includegraphics{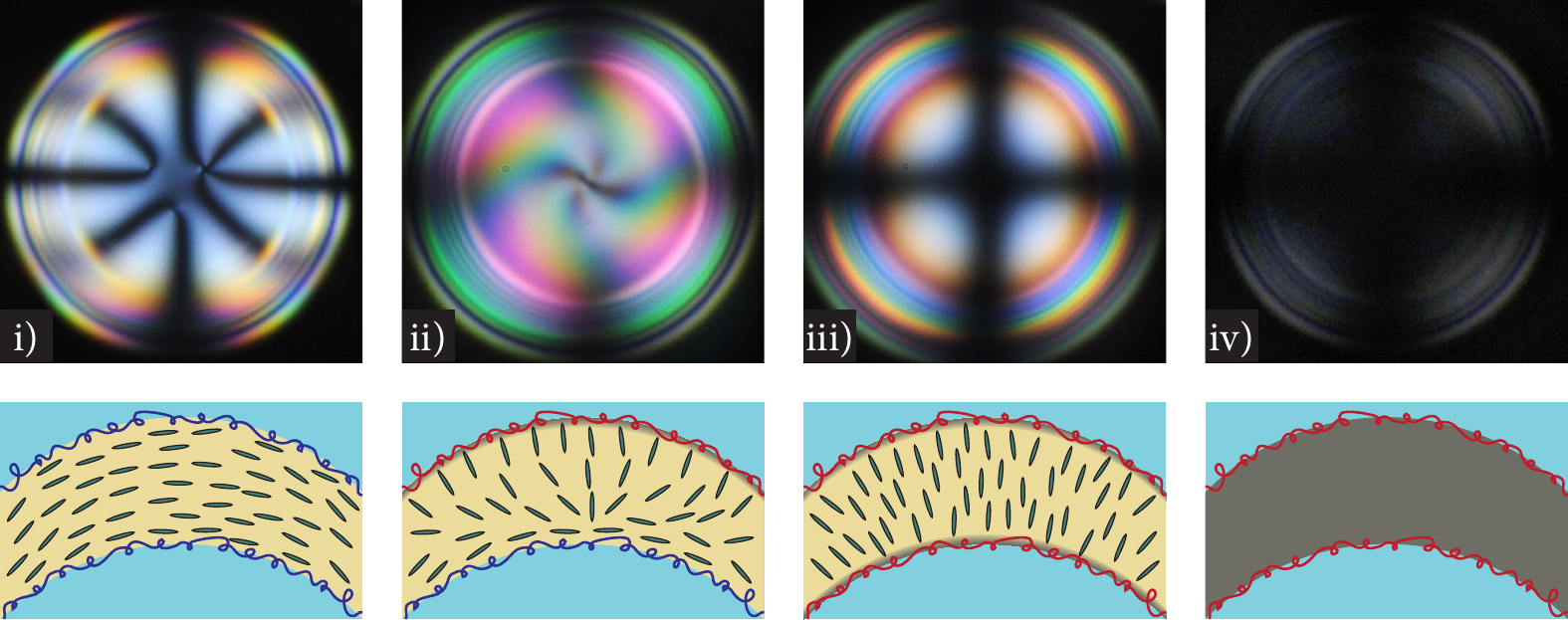}}
\caption{\label{TransitionShellPVA} \textbf{5CB shell in a \SI[detect-weight]{1}{\percent} wt PVA solution undergoing a slow temperature ramp} Top: micrographs of a shell under crossed polarizers. Bottom: cross-section of the shell schematically depicting the molecular arrangement. Throughout, the shell is \SI{110}{\um} in outer diameter and has a \SI{15}{\um} average thickness. The temperature ramp has a rate of \SI{0.01}{\degreeCelsius\per\minute}. i) For most of the nematic state, the shell displays three defects, indicating that molecules within the shell are oriented parallel to the interfaces. ii) A few tenths of degrees Celsius below the phase transition temperature of the bulk 5CB, the shell transitions to a structure having only two [+1] defects, only one of them being visible here. This is a signature of hybrid anchoring, where the molecules have planar anchoring on one interface and homeotropic anchoring on the other. iii) Even closer to the phase transition temperature, the shell displays a classic Maltese cross, which shows that the anchoring has switched to being perpendicular on both interfaces. iv) In the end, the entire shell turns isotropic. These four micrographs, taken respectively at \SI{-0.43}{\degreeCelsius}, \SI{-0.18}{\degreeCelsius}, \SI{-0.03}{\degreeCelsius} and \SI{+0.05}{\degreeCelsius} of the first apparition of bulk isotropic phase, are close-ups of frames of the timelapse shown in Supplementary Movie 5.}
\end{figure*}

This new mechanism, characterized in thin nematic films, can be profitably applied in liquid crystal shells, altering the topological constraints they are subjected to. Shells were prepared with planar anchoring at both of their interfaces. As exemplified by the lines of latitude or longitude on a globe, nematic ordering under planar anchoring is frustrated on a sphere, and has to incorporate topologically-protected defects in the director. As a consequcence of the Poincar\'{e}-Hopf theorem\cite{Poincare_theorem, Hopf_theorem}, the topological charges of the defects on such a sphere have to add up to [+2]. In nematics, topological charges higher than [+1] are unstable, and pairs of defects with opposite topological charges annihilate. Therefore, planar nematic shells can either be bivalent ($2\times$[+1]), trivalent ($1\times$[+1] + $2\times$[+1/2]) or tetravalent ($4\times$[+1/2]). Yet a three-dimensional shell is not a two-dimensional sphere: the defects are not surface points but disclination lines (singular or escaped) that span the shell thickness. Their equilibrium positions result from the balance between the repulsion of positively-charged defects and the tendency of the lines to minimize their length by migrating to the region where the shell is thinner. \cite{Alberto_FirstShell}

Liquid crystal shells are usually stabilized by adding PVA to the aqueous phase; however, those shells can be also produced without adding any stabilizer. While a regular water-in-oil-in-water double emulsion destabilizes in a matter of seconds without a surfactant, this is not the case when the middle phase is a liquid crystal, since rupturing the shell would trigger the nucleation of additional topological defects. This process has a large energy barrier associated to it, which stabilizes the double emulsion\cite{Lavrentovich_WordsAndWorlds}. For this reason, we were able to produce 5CB shells with pure water as the inner and outer solutions. These shells were subjected to a \SI{0.01}{\degreeCelsius\per\minute} temperature ramp, as shown in Supplementary Movie 4 and Fig.~\ref{TransitionsWithoutPVA}B. We witnessed the nucleation and growth of the isotropic phase within the planar nematic phase, until it took over the whole shell, the planar texture remaining unchanged throughout the entire process. 

However, when the shells were produced with \SI{1}{\percent} wt PVA in water as the inner and outer solutions, under the same experimental conditions, two anchoring transitions occurred ahead of the phase transition, as depicted in Supplementary Movie 5 and Fig.~\ref{TransitionShellPVA}. As the temperature slowly rose, three states could be unambiguously identified -- a planar state (Fig.~\ref{TransitionShellPVA}-i), a hybrid state (Fig.~\ref{TransitionShellPVA}-ii), and a radial state (Fig.~\ref{TransitionShellPVA}-iii) -- before the shell turned isotropic (Fig.~\ref{TransitionShellPVA}-iv). The planar state is the one that was discussed above; the boundary conditions are planar at both interfaces. The hybrid state corresponds to a situation where the anchoring cues are planar at one interface and non-planar at the other one. Topological defects are not required on the surface if it has homeotropic anchoring. As the director has to adapt smoothly to the conflicting planar boundary condition on the other surface, it is necessarily tilted in the bulk. In this situation, half-integer defects are no longer allowed: while the orientation of rod-like molecules lying in a plane is unchanged by a $\pi$ rotation, this is no longer the case when the molecules acquire polarity by tilting out of the plane. For this reason, hybrid shells can only display surface defects with integer topological charge, and are necessarily bivalent. The two defects are diametrically opposed to one another, as they only feel the inter-defect repulsion. These structural features, along with a characteristic vividly-colored and twisted texture, enable one to recognize hybrid shells under the microscope \cite{Teresa_TopologicalTransformations, Teresa_DefectCoalescence}. Lastly, in the radial state, shells are defect-free, with a uniform, radially-oriented director. Their optical signature is a series of colored rings barred with a large Maltese cross. It is in fact the same pattern obtained by illuminating a flat nematic cell with a uniform, vertically-oriented director in convergent light -- known as a uniaxial conoscopic interference figure \cite{Teresa_Review}.

These results show that temperature-driven anchoring transitions enable the rapid transformation between shell states with matching anchoring at both interfaces. This is a clear improvement from traditional surfactant-driven anchoring transitions, which are limited to transformations between planar to hybrid or hybrid to homeotropic shell states. On the other hand, while in suspended films the anchoring transition occurred simultaneously at the top and bottom interfaces, this is no longer the case in shells: the transition at the inner interface is slightly delayed compared to the one at the outer interface, leading to stable hybrid shells. This last result was unexpected: we hypothesize that the convex and concave natures of the outer and inner interfaces, respectively, play a role in shifting the interfacial melting temperature of the 5CB/PVA mixture. Additionally, another difference between those two systems is the absence of a stable coexistence state between the isotropic and the nematic phase: the islands from the flat films were not observed in shells.

\begin{figure*}[p!]
\centerline{\includegraphics{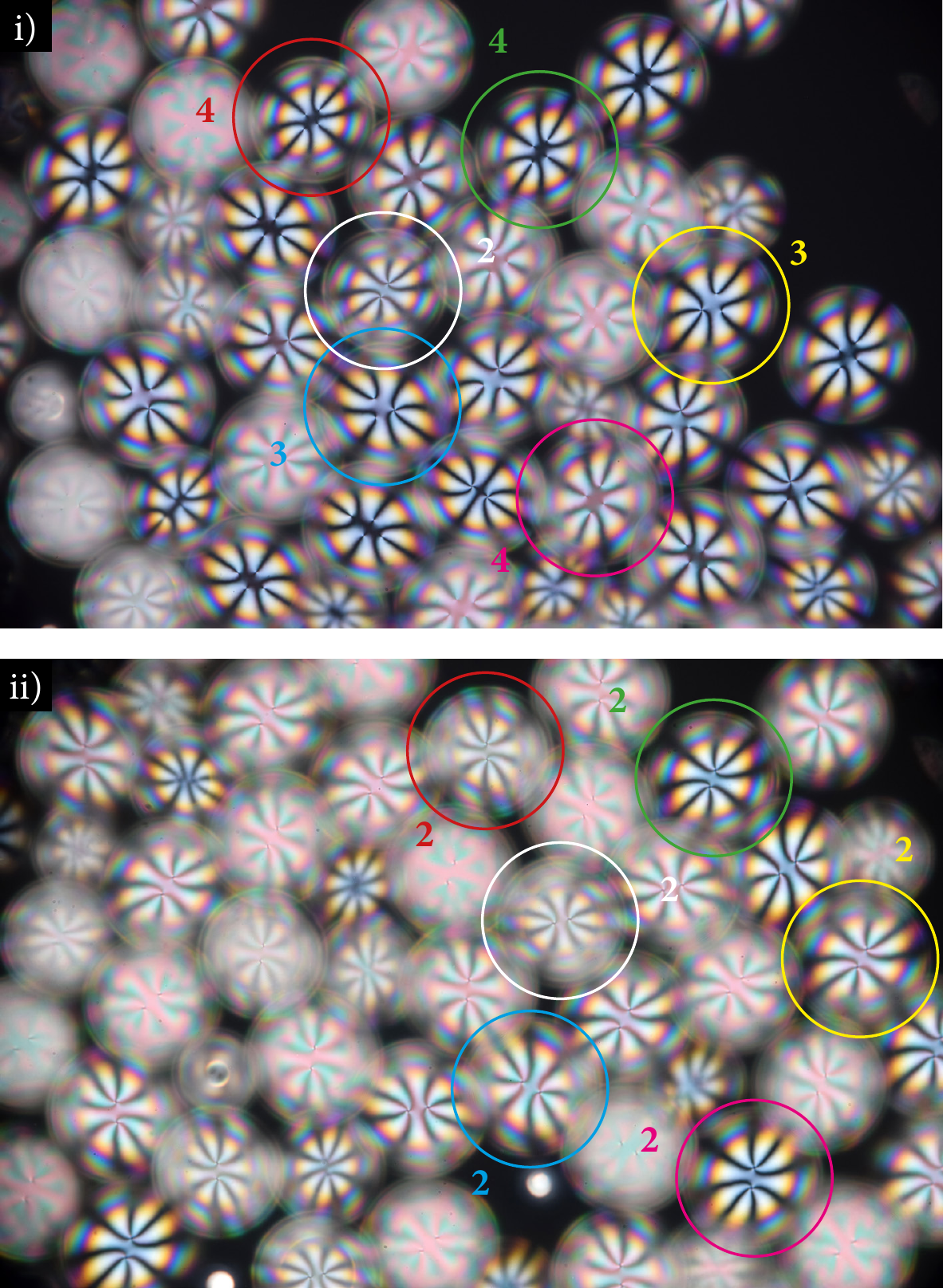}}
\caption{\label{ValencyControl} \textbf{Controlling shell valency by going through the hybrid state} i) Group of polydisperse nematic shells, i.e. displaying all three possible planar defect configurations. ii) Same group of nematic shells, showing only bivalent configurations after being subjected to a \SI{0.01}{\degreeCelsius\per\min} temperature ramp up to the isotropic phase, then to a \SI{-0.01}{\degreeCelsius\per\min} temperature ramp back to the planar phase. Residual flow within the observation cell shifted the shells' position during the temperature ramps. Those two micrographs are frames of the timelapse provided in Supplementary Movie 6.}
\end{figure*}

This new result enabled our new technique to be used with a new goal in mind: controlling the valency of a population of shells. Typically, shells are produced in an out-of-equilibrium fashion, often yielding a polydisperse population -- i.e. composed of shells with various metastable defect configurations, or valencies -- such as depicted in Fig.~\ref{ValencyControl}-i. This particular collection of shells was subjected to a \SI{0.01}{\degreeCelsius\per\minute} temperature ramp: the shells went through all of the previously described states, until all of them turned isotropic. The sample was then cooled down with a \SI{-0.01}{\degreeCelsius\per\minute} temperature ramp. The shells went through all of the states again, this time in reversed order, until the planar texture was retrieved. However, at this stage, all of the shells had the same defect configuration, as seen in Fig.~\ref{ValencyControl}-ii. This evolution can additionally be witnessed in Supplementary Movie 6.

Instead of exhibiting the lowest-energy configuration, with four [+1/2] defects \cite{Seyednejad_NemShellsSims, Koning_SphericalNematics, dePablo_MesoscaleCholShells}, they displayed a higher-energy configuration, with two [+1] defects. Indeed, our study shows that the phase transition is accompanied by an anchoring transition, which alters the topological constraints on the shell. As the hybrid state forbids the presence of half-integer defects, the shell is topologically prevented from reaching its equilibrium state. Therefore, our method turns an initially polydisperse shell population into a monodisperse population of bivalent shells. This new ability to uniformize the defect configurations of a population of planar shells could prove very useful in the perspective of shell self-assembly through defect functionalization -- a long-term goal which has motivated the experimental investigation of nematic shells since the very beginning.\cite{Nelson_TetravalentColloids} 

Temperature-induced anchoring transitions in shells have been the object of a growing interest in the past few years \cite{Lagerwall_InterfacesShells, Lisa_ChangeStripes, Lagerwall_KrafftTransition, Lagerwall_BlockCopolymers}. In particular, we note that in \cite{Lagerwall_InterfacesShells}, Lagerwall et al. performed a similar experiment on planar shells of 5CB, dispersed in \SI{1}{\percent} wt PVA in a 50:50 volume ratio water/glycerol mixture. However, no anchoring transition was reported when the shells were subjected to a \SI{0.01}{\degreeCelsius\per\min} temperature ramp. We were able to attribute experimentally the difference in behavior compared to our results to the large concentration of glycerol in the water phases, which is known to affect the surface activity of surfactant molecules.

\subsection{Conclusion}

We report a temperature-driven method for inducing molecular reorientation in a liquid crystal film, which is immediate, spatially homogeneous, and can affect both interfaces of the film. It is achieved with a precise control over the temperature of an aqueous interface of 5CB in the vicinity of its clearing point. The anchoring transition takes place if PVA is dissolved in water in concentrations greater than \SI{0.1}{\percent} in weight. We explain this behavior by the melting of the disordered PVA/5CB mixture at the interfaces of the shell, which occurs at a lower critical temperature than that of pure 5CB. The presence of this new isotropic interfacial PVA-5CB mixture could be responsible for the reorientation of the underlying pure nematic 5CB. This new method allows us to control the boundary conditions of the most widely researched shell system, shells of 5CB in aqueous solutions of PVA, without changing the composition of the system, just by using an external stimulus. It remarkably enables rapid transformations between planar and radial shell states. We find that the two interfaces in a nematic shell are not affected at exactly the same temperature: this yields a hybrid transitory state that can be used to transform an initially polydisperse population of shells, i.e. displaying all possible defect configurations, into a monodisperse population of bivalent shells. 

This work raises a number of new questions beyond its scope, which would constitute an interesting follow-up. In particular, quantitative measurements of the anchoring strength \cite{Tsukahara_AnchoringMeasurement} and precise imaging of the director throughout the bulk of the films \cite{Smalyukh_FastFCPM} at various stages of the transition would be of great interest. Our study expands the number of available parameters one can tune to trigger transitions in shells of complex liquid crystal phases. \cite{Lisa_ChangeStripes}

\begin{acknowledgement}

The authors acknowledge Christophe Blanc and Reviewer 1 for valuable feedback that helped us narrow down the nature of the mechanism at play in this phenomenon; Lisa Tran and the rest of Randall Kamien's group for experimental support and fruitful discussions; Kunyun He, Ashley Guo and Martina Clairand for experimental support; Bohdan Senyuk and Ivan Smalyukh for providing BTBP; Ye Zhou for useful discussions. This work was supported by the Agence Nationale de la Recherche (ANR) Grant No. 13-JS08-0006-01.

\end{acknowledgement}

\begin{suppinfo}

The following data is available free of charge.
\begin{itemize}
\item Temperature ramp on a homeotropic 5CB cell with solid boundaries.
\item Temperature-driven anchoring transition of a 5CB film in a \SI{0.1}{\percent} PVA solution in water (Supplementary Movie 1).
\item Temperature ramp on a homeotropic 5CB cell with solid boundaries (Supplementary Movie 2). 
\item Phase transition of a 5CB film in a solution with less than \SI{0.1}{\percent} PVA in water (Supplementary Movie 3). 
\item Phase transition of a 5CB shell without PVA in the inner and outer water phases (Supplementary Movie 4).
\item Temperature-driven anchoring transition of typical 5CB shells, with \SI{1}{\percent} PVA in the outer and inner phases (Supplementary Movie 5).
\item Temperature-driven anchoring transitions of typical 5CB shells, with \SI{1}{\percent} PVA in the water phases as the sample is heated to isotropic then cooled back to nematic (Supplementary Movie 6).
\item 5CB films doped with BTBP in a \SI{0.1}{\percent} wt PVA solution undergoing a slow temperature ramp (Supplementary Figure 1). 
\item Additional clues pertaining to the isotropic islands (Supplementary Figure 2).
\end{itemize}

\end{suppinfo}

\bibliography{AnchoringTransitions}

\end{document}


\maketitle

\section*{Temperature ramp on a homeotropic 5CB cell with solid boundaries}

A small drop of 5CB was deposited between two glass slides treated with OTS so as to impose a strong perpendicular alignment of the molecules. Under crossed polarizers, a uniformly dark area was selected, and the sample was submitted to a \SI{0.01}{\degreeCelsius\per\minute} temperature ramp. In the nematic state, conoscopy yielded a Maltese cross characteristic of homeotropic anchoring. A few tenths of degrees Celsius below the clearing point, we witnessed the rise of birefringence colors, signalling the presence of the isotropic/nematic interface. The colors and the way they nucleated were very reminiscent of our isotropic islands. However, while the island were nicely rounded by surface tension in the case of an all-fluid system, they were of a very irregular shape in the homeotropic cell, probably due to pinning on surface defects. Then fully black patches started appearing where the isotropic phase had reached both walls of the cell, until they took over the entire cell. This experiment in a more well-known situation validates the weakly out-of-plane nature of the anchoring between 5CB and its isotropic phase.

\section*{Supplementary movies}

\begin{itemize}
\item SupMov1\_HeatingFilmPVA: timelapse of the temperature-driven anchoring transition of a 5CB film in a \SI{0.1}{\percent} PVA solution in water, corresponding to Fig.~2.
\item SupMov2\_HeatingHomeoCell: timelapse of a 5CB homeotropic cell heated quasi-statically in the vicinity of its clearing point.
\item SupMov3\_HeatingFilmPureWater: timelapse of the phase transition of a 5CB film in a solution with less than \SI{0.1}{\percent} PVA in water, corresponding to Fig.~4A. 
\item SupMov4\_HeatingShellPureWater: real-time movie of the phase transition of a 5CB shell without PVA in the inner and outer water phases, corresponding to Fig.~4B.
\item SupMov5\_HeatingShellPVA: timelapse of the temperature-driven anchoring transition of typical 5CB shells, with \SI{1}{\percent} PVA in the outer and inner phases, corresponding to Fig.~5.
\item SupMov6\_HeatingCoolingShellsPVA: timelapse of the temperature-driven anchoring transitions of typical 5CB shells with \SI{1}{\percent} PVA in the water phases as the sample is heated to isotropic then cooled back to nematic, corresponding to Fig.~6.
\end{itemize}

\section*{Supplementary figures}

\begin{figure}[ht!]
\centerline{\includegraphics[scale=.9]{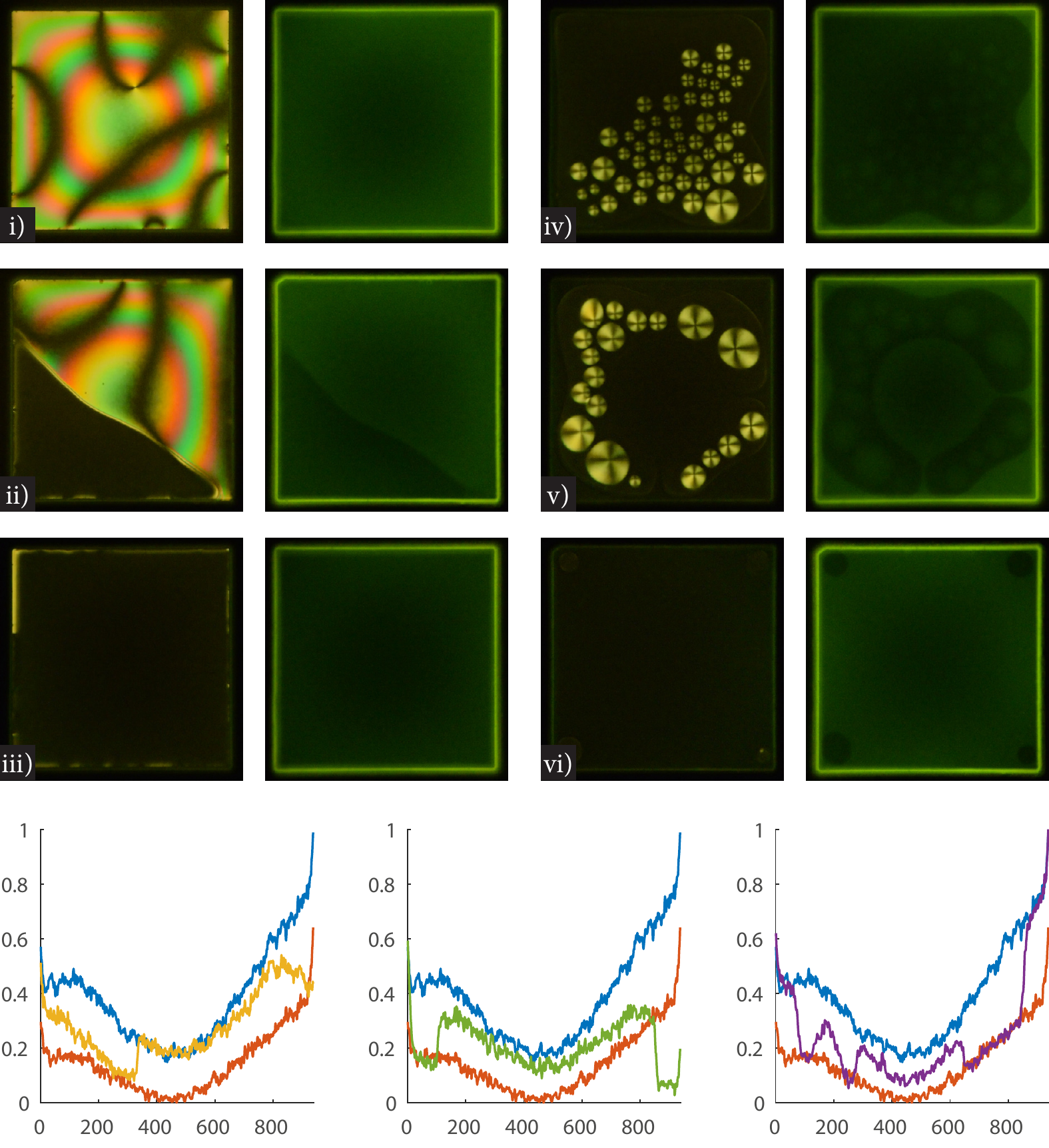}}
\caption{\label{FluorescentFilm} \textbf{5CB films doped with BTBP in a \SI{0.1}{\percent} wt PVA solution undergoing a slow temperature ramp.} Left: micrographs between crossed polarizers, lit under both white light and near-ultraviolet light. Right: micrographs in fluorescence mode (single polarizer), lit under near-ultraviolet light. Bottom: average intensity profiles in a 25-pixel wide band along the SW-NE diagonal of the films. The U-shape of all the graphs is due to the concavity of the film: there is more material on the edges, hence more dopant, on the edges than in the center. Throughout, the films are \SI{430}{\um} wide, about \SI{10}{\um} thick close to the edges, about \SI{5}{\um} thick in the center, and fixed on a suspended TEM grid. The temperature ramp has a rate of \SI{0.01}{\degreeCelsius\per\minute}. The series of pictures depicts typical stages of the evolution of a film, as temperature is increased: i) planar state (blue curve), ii) transition from the planar to the homeotropic state (yellow curve), iii) homeotropic state (red curve), iv) early stage with the birefringent islands, v) late stage with the birefringent islands (purple curve), vi) almost all of the film has turned isotropic (green curve). The transition between planar and homeotropic levels shows a clear jump in fluorescence intensity. The isotropic profile shows intermediate values, about two thirds of the way between planar and homeotropic levels. The corners appear very dark due to remaining patches of homeotropic nematic phase. The late island stage profile shows a similar behavior as the isotropic profile; the peak on the left part corresponds to a birefringent island.} 
\end{figure}

\begin{figure}[ht!]
\centerline{\includegraphics{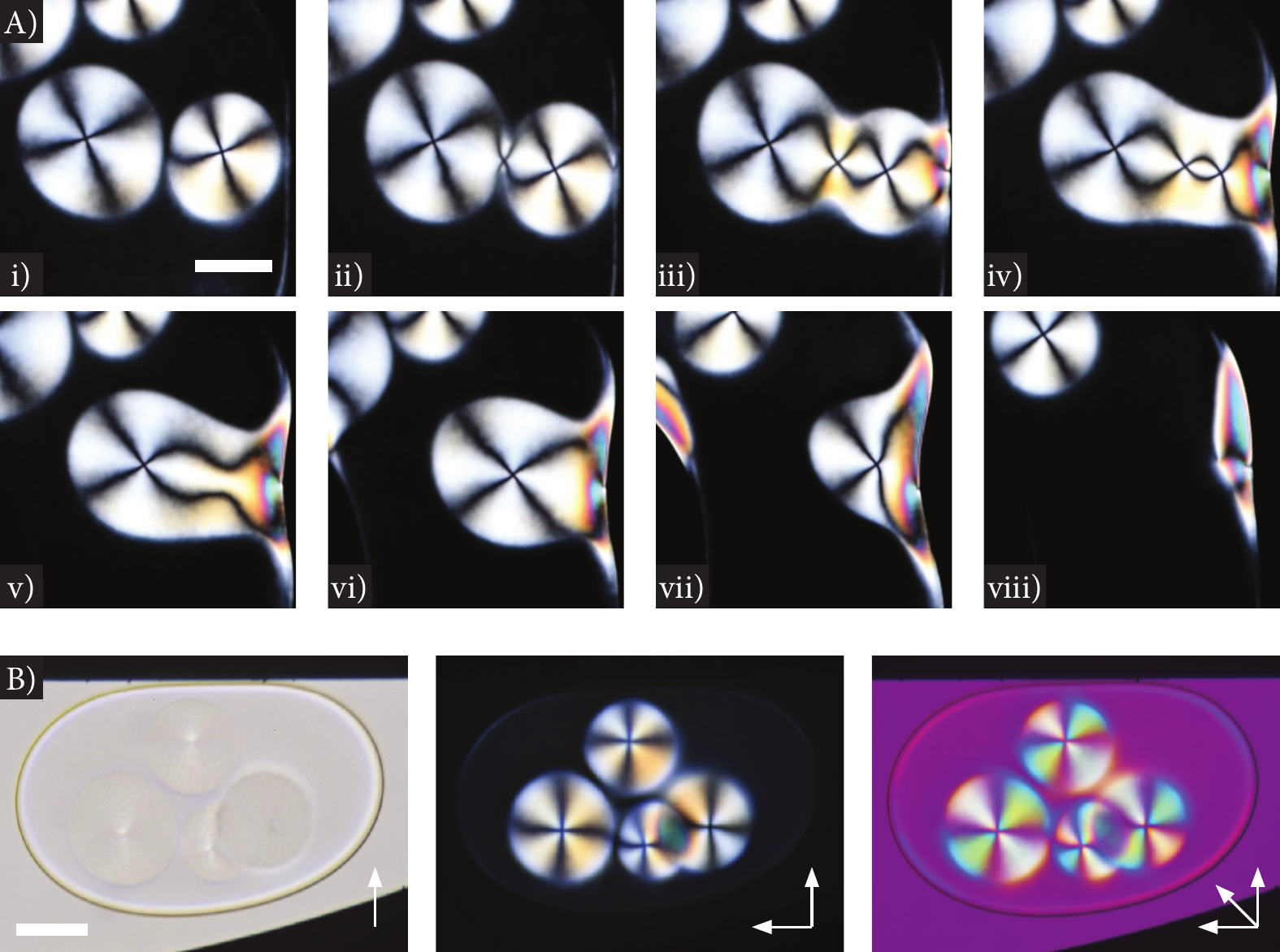}}
\caption{\label{AdditionalIslands} \textbf{Additional clues pertaining to the isotropic islands.} Micrographs showing isotropic islands on a 5CB film in a \SI{0.1}{\percent} wt PVA solution in water, scale bars 50 $\mu$m. A) Coalescence of islands and merging into the isotropic phase. Time series as the temperature is linearly increased with a \SI{0.01}{\degreeCelsius\per\minute} temperature ramp. Two isotropic islands are initially apart (i), they coalesce, creating a [-1] defect at the neck of the junction (ii-iii). The new defect then annihilates with one of the two [+1] initial defects (iv-v). Lastly, the remaining island fuses with the isotropic phase, growing from the right side of the image (vi-viii). B) Overlapping islands. Pictures in bright field, crossed polarizers, and crossed polarizers with first-order retardation plate. The temperature is a few tenths of degrees Celsius below the clearing point of pure 5CB. One island is located on one of the two interfaces of the film, while the other three islands are located on the second interface. }
\end{figure}